\documentclass{WileyMSP-template}

\usepackage{amsmath}
\usepackage{siunitx}
\usepackage{mathtools}
\usepackage{physics}
\usepackage{amssymb}
\usepackage{soul}
\usepackage{jabbrv}
\DeclareMathOperator*{\argmin}{arg\,min}
\begin{document}

\pagestyle{fancy}
\rhead{\includegraphics[width=2.5cm]{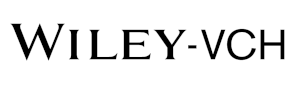}}

\title{End-to-end metasurface design for temperature imaging via broadband Planck-radiation regression}

\maketitle

% Author: Please give full first and last names for authors and include * after the name of all corresponding authors

\author{Sophie Fisher*}

\author{Gaurav Arya}

\author{Arka Majumdar}

\author{Zin Lin}

\author{Steven G. Johnson}

% Dedication

\dedication{}

% Affiliations: Please provide adacemic titles (Prof. or Dr.) for all authors where applicable, and include an institutional email address for all corresponding authors
\begin{affiliations}
S. Fisher\\
Department of Electrical Engineering and Computer Science \\ Massachusetts Institute of Technology \\
Cambridge, MA 02139, USA \\
E-mail: sefisher@mit.edu \\ 

G. Arya, S. G. Johnson \\
Department of Mathematics \\
Massachusetts Institute of Technology \\
Cambridge, MA 02139, USA \\

A. Majumdar \\
Department of Electrical and Computer Engineering \\
University of Washington \\
Seattle, WA 98195, USA \\

A. Majumdar \\
Department of Physics \\
University of Washington \\
Seattle, WA 98195, USA \\

Z. Lin \\
Bradley Department of Electrical and Computer Engineering \\
Virginia Tech \\
Blacksburg, VA 24060, USA \\

\end{affiliations}

% Keywords: Please provide a minimum of three and a maximum of seven keywords, separated by commas


% Abstract should be written in the present tense and impersonal style (i.e., avoid we), and be at most 200 words long
\begin{abstract}
We present a theoretical framework for temperature imaging from long-wavelength infrared thermal radiation (e.g.~\SIrange[range-phrase=\textendash, range-units=single]{8}{12}{\micro\metre}) through the end-to-end design of a metasurface-optics frontend and a computational-reconstruction backend. We introduce a new nonlinear reconstruction algorithm, ``Planck regression,” that reconstructs the temperature map from a grayscale sensor image, even in the presence of severe chromatic aberration, by exploiting blackbody and optical physics particular to thermal imaging. We combine this algorithm with an end-to-end approach that optimizes a manufacturable, single-layer metasurface to yield the most accurate reconstruction. Our designs demonstrate  high-quality, noise-robust reconstructions of arbitrary temperature maps (including completely random images) in simulations of an ultra-compact thermal-imaging device. We also show that Planck regression is much more generalizable to arbitrary images than a straightforward neural-network reconstruction, which requires a large training set of domain-specific images.
\end{abstract}

% Text: Please use section headings and subheadings as specified below. For communications, all section headings apart from Experimental Section should be removed
% Please make the first reference to a display item bold: \textbf{Figure 1}
% Do not abbreviate Figure, Equation, etc.; display items are always singular, i.e., Figure 1 and 2.
% Equations are always singular, i.e., Equation 1 and 2, and should be inserted using the {equation} environment, not as graphics
% Please do not use footnotes in the text, additional information can be added to the Reference list.

\section{Introduction}

In this paper, we present a new nonlinear reconstruction algorithm for thermal imaging, ``Planck regression,'' based on the physics of blackbody radiation---inferring the temperature map $\vb{T}(x,y)$ from a grayscale image of broadband long-wavelength infrared (LWIR) thermal emission \cite{vollmerInfraredThermalImaging2017}. We combine this with fullwave end-to-end co-design~~\cite{aryaEndtoEndOptimizationMetasurfaces2024, liTranscendingShiftinvarianceParaxial2023, linEndtoendMetasurfaceInverse2022, linEndtoendNanophotonicInverse2021,burgosDesignFrameworkMetasurface2021, tsengNeuralNanoopticsHighquality2021} of a manufacturable single-layer metasurface~\cite{khorasaninejadMetalensesVisibleWavelengths2016, engelbergAdvantagesMetalensesDiffractive2020, chenBroadbandAchromaticMetalens2018, khorasaninejadVisibleWavelengthPlanar2017, pestourieInverseDesignLargearea2018, yuFlatOpticsDesigner2014}, yielding a compact thermal-imaging system that reconstructs arbitrary (even random) objects with robust performance ($2 \%$ error with $4 \%$ noise). Our Planck regression algorithm (Section~\ref{section_theory}) exploits a strong physical-information prior particular to the thermal regime: namely, that the object is a black body whose emitted radiation spectrum follows Planck's law~\eqref{planckslaw}. Under this assumption, the reconstruction can be formulated as a \emph{nonlinear} least-squares problem that de-convolves the noise-corrupted sensor image using the known point spread functions (PSFs) of the system while directly estimating the temperature map of the thermal object. Planck regression is quite different from earlier approaches that did not exploit blackbody physics~\cite{aryaEndtoEndOptimizationMetasurfaces2024, liTranscendingShiftinvarianceParaxial2023, linEndtoendMetasurfaceInverse2022, linEndtoendNanophotonicInverse2021,burgosDesignFrameworkMetasurface2021, tsengNeuralNanoopticsHighquality2021, baekSingleshotHyperspectralDepthImaging2021, sitzmannEndtoendOptimizationOptics2018}, discussed further below. Our algorithm is effective even without any end-to-end design: when combined with ordinary metalenses~\cite{khorasaninejadMetalensesVisibleWavelengths2016, khorasaninejadVisibleWavelengthPlanar2017, pestourieInverseDesignLargearea2018} (designed \emph{independently} of the reconstruction) that suffer from severe chromatic aberration, our method nevertheless yields accurate reconstructions ($8 \%$ error with $4 \%$ noise in Section~\ref{section_results}). Our algorithm assumes no inherent pattern in the image and even works when every pixel is an independent random temperature, whereas we show for comparison that a convolutional neural network (CNN) can only be trained to reconstruct a low-dimensional subset of images (e.g.~pictures of circles). Furthermore, we demonstrate even better performance by applying the paradigm of end-to-end design~\cite{aryaEndtoEndOptimizationMetasurfaces2024, liTranscendingShiftinvarianceParaxial2023, linEndtoendMetasurfaceInverse2022, linEndtoendNanophotonicInverse2021,burgosDesignFrameworkMetasurface2021, tsengNeuralNanoopticsHighquality2021, baekSingleshotHyperspectralDepthImaging2021, sitzmannEndtoendOptimizationOptics2018}: we \emph{co-optimize} the optics (a metasurface of millions of subwavelength Si pillars) along with the reconstruction algorithm to  minimize the overall reconstruction error. That is, we let optimization discover the physically-realizable PSFs that are best suited to Planck regression (Section~\ref{section_theory}). This reduces the reconstruction error by a factor of four (Section~\ref{section_results}).  Despite the nonconvexity of the end-to-end design problem, we obtain similar performance (albeit with different designs) from a variety of starting metasurfaces; similarly, we find that Planck regression empirically converges to an accurate temperature map regardless of its starting point. 

Conventional thermal imaging centers on the ideal of a broadband achromatic lens: if the \emph{total} radiation from one point in a thermal source is focused onto a single sensor pixel, then the integrated Planck's law implies a one-to-one relationship between intensity and temperature, which can be determined with the help of a device calibration~\cite{vollmerInfraredThermalImaging2017}. This assumes that the emitter is approximately a black body, because many materials at LWIR wavelengths empirically have emissivity $> 90\%$~\cite{steketeeSpectralEmissivitySkin1973, avdelidisEmissivityConsiderationsBuilding2003, barnesTotalEmissivityVarious1947}. However, obtaining a broadband LWIR lens is challenging, typically requiring bulky optics. Compact lenses, especially ultra-thin diffractive or metasurface lenses, suffer from chromatic aberrations \cite{fanHighNumericalAperture2018, huangLongWavelengthInfrared2021, kignerMonolithicAllsiliconFlat2021, linWideFieldofViewLargeAreaLongWave2024, meemBroadbandLightweightFlat2019, nalbantTransmissionOptimizedLWIR2022, wirth-singhLargeFieldofviewThermal2023, presuttiFocusingBandwidthAchromatic2020}, which one then attempts to correct via computational post-processing using a variety of algorithms~\cite{heSingleimagebasedNonuniformityCorrection2018, huangBroadbandThermalImaging2024, liuShutterlessNonuniformityCorrection2018, saragadamThermalImageProcessing2021, saragadamFoveatedThermalComputational2024, hardieMAPEstimatorSimultaneous2007, hardieScenebasedNonuniformityCorrection2000}.  However, these aberration-correcting algorithms are still targeting the ideal of an achromatic-lens intensity image by taking into account the \emph{optical} physics, often supplemented by NN regularizations~\cite{ulyanovDeepImagePrior2020, saragadamThermalImageProcessing2021, saragadamFoveatedThermalComputational2024}, without including the \emph{thermal} physics.

A ``metasurface'' is a subwavelength-patterned surface designed to shape free-space radiation~\cite{yuFlatOpticsDesigner2014, chenBroadbandAchromaticMetalens2018, engelbergAdvantagesMetalensesDiffractive2020, khorasaninejadMetalensesVisibleWavelengths2016, khorasaninejadVisibleWavelengthPlanar2017, pestourieInverseDesignLargearea2018}. The subwavelength-scale patterning means that the scattering of light through a metasurface involves fullwave electromagnetic effects, such as frequency- and polarization-sensitive resonances, as opposed to traditional diffractive optics~\cite{engelbergAdvantagesMetalensesDiffractive2020} in which scattering can be locally described by ray optics because the pattern varies much slower than the wavelength. Metasurfaces have been designed and fabricated for various applications at many wavelengths, most commonly by lithographic patterning of dielectric thin films, e.g.~into arrays of subwavelength pillars~\cite{khorasaninejadMetalensesVisibleWavelengths2016} or holes~\cite{limHighAspectRatio2021} of varying diameters or shapes.  In Section~\ref{section_results}, we show theoretical results based on silicon-pillar metasurfaces similar to those fabricated for LWIR applications in previous work~\cite{fanHighNumericalAperture2018, huangLongWavelengthInfrared2021, kignerMonolithicAllsiliconFlat2021, linWideFieldofViewLargeAreaLongWave2024, meemBroadbandLightweightFlat2019, wirth-singhLargeFieldofviewThermal2023}.

The paradigm of end-to-end optical co-design, in which the optics is designed to directly minimize the inference/reconstruction error of a post-processing algorithm, has recently been applied to a variety of problems, including full-color visible imaging~\cite{tsengNeuralNanoopticsHighquality2021, sitzmannEndtoendOptimizationOptics2018}, compressed sensing~\cite{aryaEndtoEndOptimizationMetasurfaces2024}, and hyperspectral imaging~\cite{linEndtoendMetasurfaceInverse2022, baekSingleshotHyperspectralDepthImaging2021, linEndtoendNanophotonicInverse2021}. In principle, a hyperspectral imaging system could be used for thermal imaging: given the intensity at many different wavelengths, one could fit to Planck's law and determine temperature and even perhaps emissivity spectra.  However, this requires many more camera pixels to reconstruct a vast amount of intermediate information (images at many wavelengths), whereas by directly reconstructing the temperature map we can obtain an image at 50\% of the resolution of the raw sensor image (Section~\ref{section_results}).

\section{Theory: Image formation and reconstruction}\label{section_theory}

\begin{figure*}
    \centering
    \includegraphics[width=\linewidth]{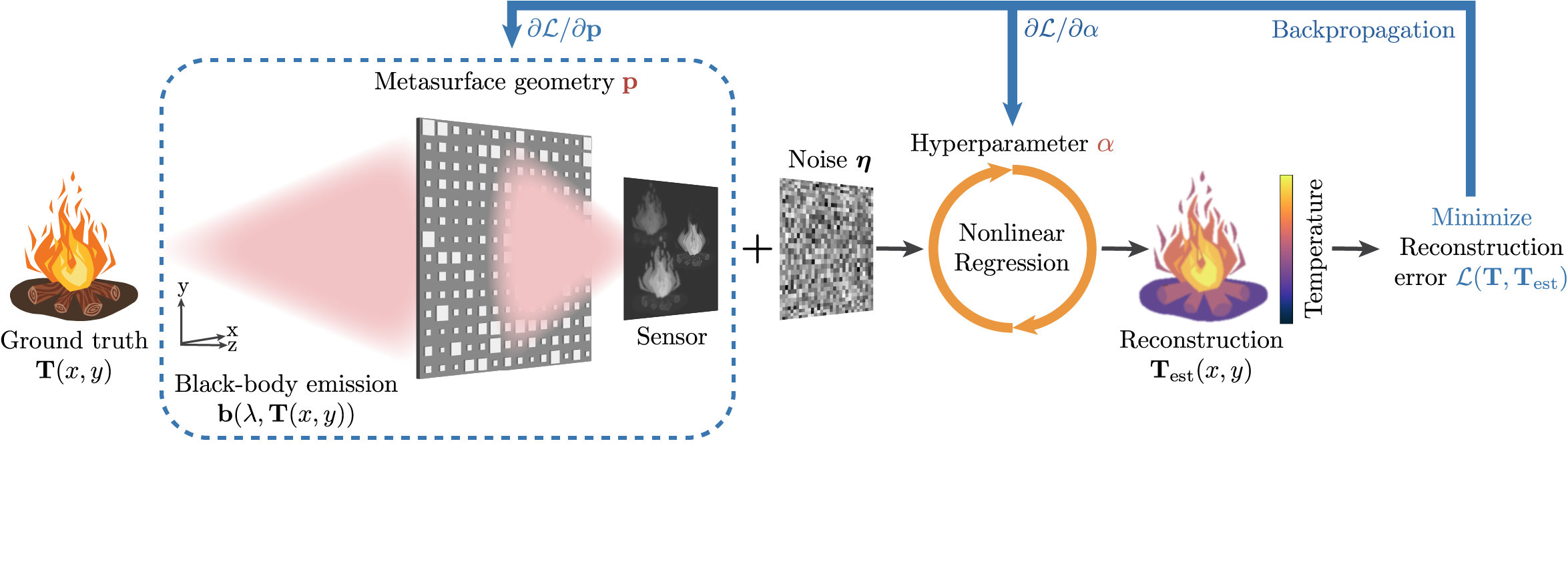}
    \caption{The end-to-end optimization pipeline. The differentiable forward pipeline includes a physical model of the monochrome sensor image formed from the blackbody emitting object with temperature map $\vb{T}$, together with a computational reconstruction of the object’s temperature map $\vb{T}_{\text{est}}$. The reconstruction error $\mathcal{L}(\vb{T}, \vb{T}_{\text{est}})$ is formed at the end of the forward pipeline. The optimization minimizes this error with respect to the metasurface parameters $\vb{p}$ and the reconstruction hyper-parameter $\alpha$ by back-propagating the gradients through the image formation and reconstruction algorithm of the forward pipeline. }
    \label{endtoendfigure}
\end{figure*}
Here we outline the theory of the image reconstruction and the end-to-end optimization problem (Figure~\ref{endtoendfigure}), consisting of a differentiable forward pipeline that models the thermal image formation and computes the nonlinear Planck regression, and a backward pipeline that backpropagates the gradients of the reconstruction error with respect to optical/reconstruction parameters (necessary for optimizing the optical design) through the forward pipeline.

We model the ground truth object as a 2D grid $(x,y)$ of thermally emitting point sources at a fixed distance $(z)$ from the metasurface. We assume that each point source is nearly a black body within the LWIR bandwidth with emissivity $\varepsilon_{f}(x,y) \approx 1$ and temperature $T(x,y)$ (As discussed in Section~\ref{section_results}, an emissivity differing significantly from~$1$ yields only moderate errors in reconstructed temperature.) The emitted light intensity (power per area, frequency and solid angle) of each point source is therefore given by Planck's law~\cite{vollmerInfraredThermalImaging2017}
\begin{equation}
    b(f, T) = \frac{2 h f^3}{c^2} \frac{1}{\exp{ \big(\frac{h f}{k_B T}\big)} - 1} \, ,
    \label{planckslaw}
\end{equation}
where $b(f, T)$ is the ``spectral radiance'' at a temperature $T$ and frequency $f$, $k_B$ is the Boltzmann constant, $h$ is the Planck constant, and $c$ is the speed of light in vacuum. 

The thermal emission is scattered by the metasurface onto a thermal sensor such as a microbolometer~\cite{vollmerInfraredThermalImaging2017}, which produces a grayscale, noise-corrupted image. This image formation at each frequency can be modeled as a convolution of the object's intensity with the point-spread function (PSF) of the optics~\cite{goodmanIntroductionFourierOptics2005}, where the PSF is defined as the intensity image produced by a monochromatic normally incident plane wave (which can be interpreted as the light of a point source far away~\cite{goodmanIntroductionFourierOptics2005}). As is typical for metasurface modeling, we further employ a paraxial approximation corresponding to a shift-invariant PSF~\cite{goodmanIntroductionFourierOptics2005}. Then, for temporally incoherent light (such as thermal emission), broadband image formation can be modeled as a frequency integral of the PSF convolutions over the spectral bandwidth, in our case weighted by Planck's law.  We assume a sensor that responds to a finite LWIR bandwidth $\Delta f$, e.g.~wavelengths of \SIrange[range-phrase=\textendash, range-units=single]{8}{12}{\micro\metre}~\cite{vollmerInfraredThermalImaging2017}. Let $\vb{p}$ denote the vector of geometric degrees of freedom of the optics (e.g. the metasurface pillar widths), and let $\vb{T}$ denote the vector of point-source temperatures corresponding to a flattened  (``vectorized''~\cite{golubMatrixComputations2013}) two-dimensional grid of the point sources, representing the full spatial dependence of the object. Combining the above steps, our image-formation model, which takes as input a temperature map vector $\vb{T}$ and outputs an image vector $\vb{v}$, is given by:
\begin{equation}
    \underbrace{ \vb{v} }_{\text{image}} = \int_{\Delta f} d f \,  \underbrace{ \hat{G} (f, \vb{p} )}_{\mathclap{\substack{\text{convolution} \\ \text{with PSFs}}}} \,  \underbrace{ \vb{b}(f, \vb{T}) }_{\text{emission}} + \underbrace{\vb*{\eta} }_{\text{noise}}
    \label{imageformation} \, ,
\end{equation}
where $\hat{G} (f, \vb{p} )$ is a linear operator whose action is to convolve the emission with the PSF at a given frequency, which more explicitly is given by 
\begin{equation}
   \hat{G} (f, \vb{p}) \vb{b}(f, \vb{T})|_{xy} = \iint dx' dy' \text{PSF}(f, \vb{p}, x-x', y-y') \vb{b}(f, T(x', y'))),
\end{equation}
$\vb{b}(f, \vb{T})$ is Equation~\eqref{planckslaw} applied elementwise to $\vb{T}$, and $\vb*{\eta}$ is sampled from the noise profile of the sensor. 
We model the noise as zero-mean Gaussian white noise with standard deviation $\sigma: \vb*{\eta} \propto \vb{N}(0,\sigma)$~\cite{sitzmannEndtoendOptimizationOptics2018}. (In practice, thermal sensors have more complicated noise profiles, stemming from effects such as self-heating of the sensor and stray reflections in addition to shot noise and electronic-read noise~\cite{vollmerInfraredThermalImaging2017}, whose impact could be investigated in future work.)

Given the noisy image $\vb{v}$, the goal of our image processing/reconstruction is to output a reconstructed temperature map $\vb{T}_{\text{est}}(x,y)$. We pose the reconstruction as the following nonlinear optimization problem, which we call``Planck regression,'' that tries to find a temperature map that most closely reproduces the image:
\begin{equation}
    \vb{T}_{\text{est}} = \argmin_{{ \vb*{\tau}\ge 0}} \big\Vert \vb{v} - \int_{\Delta f} d f \, \hat{G} (f, \vb{p} ) \vb{b}(f, \vb*{\tau}) \big\Vert_2^2 + \alpha \big\Vert \vb*{\tau} - T_{\text{bg}} \big\Vert_2^2 \, .
    \label{reconstruction}
\end{equation}
The second term being minimized is a Tikhonov/$L_2$ regularizaton term~\cite{tarantolaInverseProblemTheory2005}, where $\alpha$ is a non-negative hyper-parameter of the reconstruction that tunes the strength of the regularization, and $ T_{\text{bg}}$ is the expected value of the background temperature. The role of the Tikhonov regularization is to better condition the inverse problem, trading off robustness to sensor noise for accuracy in the reconstruction, and to incorporate prior knowledge of the ambient background temperature. (Other priors such as sparsity or smoothness could be imposed through similar terms~\cite{tarantolaInverseProblemTheory2005}.) Equation~\eqref{reconstruction} is distinct from previous end-to-end works, which also used deconvolution algorithms but reconstructed the intensity image rather than the temperature, often using a linear regression~\cite{aryaEndtoEndOptimizationMetasurfaces2024, liTranscendingShiftinvarianceParaxial2023, linEndtoendMetasurfaceInverse2022, linEndtoendNanophotonicInverse2021, sitzmannEndtoendOptimizationOptics2018}. In contrast, Planck regression incorporates \emph{thermal} physics [Equation~\eqref{planckslaw}] into the deconvolution, solving directly for the temperature using a \emph{nonlinear} regression (via algorithms described in Section~\ref{section_methods}).

We can finally pose an end-to-end optimization problem, to find the ``best optics'' for the Planck regression reconstruction, as:
\begin{equation}
    \min_{\vb{p},\alpha} \underbrace{\bigg\langle \frac{\| \vb{T} - \vb{T}_{\text{est}}\|_2^2 }{\| \vb{T} \|_2^2} \bigg\rangle_{ \vb{T}, \vb*{\eta} }}_{\mbox{loss function }\mathcal{L(\vb{p},\alpha)}}
    \label{endtoend_objective} \, ,
\end{equation}
where $\langle \cdot \rangle_{\vb{T}, \vb*{\eta}}$ denotes an average over a training set of temperature maps and noise distributions. That is, we minimize the average relative mean square error (MSE) of the reconstructions over a training set, with respect to both the metasurface/optics parameters $\vb{p}$ and the reconstruction hyper-parameter $\alpha$. This formulation is a \emph{bilevel} optimization problem~\cite{sinhaReviewBilevelOptimization2017}, in which the nonlinear reconstruction is nested within the end-to-end design problem.

\section{Computational methods}\label{section_methods}

In this section, we outline the computational methods that we use to solve Equations~(\ref{imageformation}--\ref{endtoend_objective}).

First, we compute the PSFs by the well-established ``locally periodic approximation'' (LPA) for metasurface design~\cite{pestourieInverseDesignLargearea2018}. This involves first transmitting a normally incident planewave (or a point source far away) through the metasurface via LPA, followed by a near-to-far field transformation~\cite{goodmanIntroductionFourierOptics2005} that propagates the fields from the metasurface to the sensor plane. 

Our metasurfaces (described in Section~\ref{section_results}) consist of millions of dielectric pillars of varying widths arranged on a periodic grid, similar to our recently reported all-silicon metalenses~\cite{huangLongWavelengthInfrared2021, huangBroadbandThermalImaging2024, saragadamFoveatedThermalComputational2024, wirth-singhLargeFieldofviewThermal2023, tsengNeuralNanoopticsHighquality2021}. LPA approximates the transmission through each unit cell (pillar) by periodic boundary conditions, which has proven to be accurate for lens-like metasurfaces where the unit cells are mostly slowly varying~\cite{pestourieInverseDesignLargearea2018, perez-arancibiaSidewaysAdiabaticityRay2018, liInverseDesignEnables2022}.  For a given periodic unit cell, the transmission is computed using rigorous coupled-wave analysis (RCWA)~\cite{liuS4FreeElectromagnetic2012}.  To further accelerate computations, we sample a range of pillar widths at each frequency and then simply interpolate to obtain the complex transmission coefficient for any unit cell~\cite{pestourieInverseDesignLargearea2018}. Here, we use Chebyshev polynomial interpolation~\cite{trefethenApproximationTheoryApproximation2013}, denoted below by~$\vb{t}(f, \vb{p})$. The transmitted electric fields are then given by $\vb{E}_{\text{transmitted}} = \vb{t}(f, \vb{p}) \odot \vb{E}_{\text{incident}}$, where $\vb{E}_{\text{incident}}$ is a vector corresponding to the metasurface unit cells that represents the electric fields at the metasurface plane produced by a point source with frequency $f$ at the object plane, $\vb{t}(f, \vb{p})$ is the polynomial ``surrogate'' model applied elementwise to each unit cell parameter, and $\odot$ denotes the elementwise product of two vectors. The full PSF is therefore given by
\begin{equation}
    \vb{PSF}(f, \vb{p}) = \mathrm{bin} \bigg( \frac{\mathrm{intensity}\left[ \text{FF}(\vb{t}(f, \vb{p}) \odot \vb{E}_{\text{incident}}) \right]}{f} \bigg) \, ,
    \label{PSFs}
\end{equation}
where FF is the near-to-far-field transformation that convolves the transmitted electric fields with the free-space near-to-far-field Green's function to obtain the electric fields at the sensor plane~\cite{goodmanIntroductionFourierOptics2005,pestourieInverseDesignLargearea2018}, $\mathrm{intensity}[\cdots]$ is the electric-field intensity at each point on the sensor, $\mathrm{bin}(\cdots)$ is a linear operator which integrates the intensity within each sensor pixel, and $\vb{PSF}$ is a vector of PSF values corresponding to a flattened two-dimensional grid of the sensor pixels. Each PSF is divided by frequency~$f$ to convert the intensity to a value that is proportional to the photon count. Here, we employ a paraxial approximation, whereby a lateral shift of the incident wave corresponds only to a shift of the image, so that the PSF is treated as a fixed pattern (independent of the incident wave position) that is convolved with the object's emission. This approximation is widely used in computational imaging where the object is sufficiently far-field, and is justified in the problems considered here where the field of view ($\pm 2 \degree$) and the NA (0.2) are small~\cite{goodmanIntroductionFourierOptics2005}.

Convolution with this PSF is represented by $\hat{G} (f, \vb{p} )$  in Equation~\eqref{imageformation}.
Since $\hat{G} (f, \vb{p} )$ is at least $10^3 \times 10^3$ in dimension, we avoid constructing the matrix directly and instead apply the convolutions implicitly through FFT operations~\cite{goodmanIntroductionFourierOptics2005, frigoDesignImplementationFFTW32005}. Since we only consider overdetermined problems where the number of sensor pixels is greater than the number of temperature map pixels, $\hat{G}$ always has more rows than it has columns. To carry out the integration over frequency~$f$ in the image formation model [Equation~\eqref{imageformation}], we use a Clenshaw--Curtis quadrature \cite{trefethenApproximationTheoryApproximation2013} which involves computing the PSFs at a set of Chebyshev points within the spectral bandwidth.

Our Planck-regression problem [Equation~\eqref{reconstruction}] can be solved efficiently using a local, gradient-based nonlinear-optimization algorithm (subject to bound constraints $\tau \ge 0$). We employ the limited-memory BFGS~\cite{nocedalUpdatingQuasiNewtonMatrices1980, liuLimitedMemoryBFGS1989} (L-BFGS) algorithm.  The algorithm requires the derivative of the objective function (the squared error + Tikhonov regularization) with respect to the optimization parameters $\tau$; it is straightforward to compute these derivatives analytically. We found that the L-BFGS optimization typically converges in $\sim 100$ iterations to an accurate reconstruction regardless of the initialization.

Finally, we formulate our end-to-end design problem as a stochastic optimization task [Equation~\eqref{endtoend_objective}] ~\cite{lanFirstorderStochasticOptimization2020}: we minimized the expected error over an infinite ``training'' set of random temperature maps, drawn from a uniform distribution between $T_{\text{low}}$ to $T_{\text{high}}$, and noise drawn from Gaussian white noise (taking the amplitude and standard deviation of the distribution to be a fixed percentage of the mean image value; see Appendix~\ref{appendix}). We solve this optimization problem using the well-known stochastic gradient-descent algorithm Adam~\cite{kingmaAdamMethodStochastic2014}, in which at each iteration we sample a random mini-batch of three temperature maps and noise profiles. To compute the gradients $\partial \mathcal{L} / \partial \vb{p}$ and  $\partial \mathcal{L} / \alpha$, we analytically backpropagate the derivatives through the reconstruction backend and the physics frontend using a custom adjoint algorithm~(Appendix~\ref{appendix}) rather than relying on automatic-differentiation libraries~\cite{innesDifferentiableProgrammingSystem2019} which we found to perform poorly for this problem (presumably because they do not exploit the full algebraic structure of this problem). Gradient calculation for bilevel optimization is achieved by implicitly differentiating the Karush-Kuhn-Tucker (KKT) conditions~\cite{boydConvexOptimization2004, blondelEfficientModularImplicit2021} associated with the reconstruction and employing an adjoint method (also called ``reverse mode'' or ``backpropagation'')~\cite{strangComputationalScienceEngineering2012}, which allows one to compute the gradient with respect to \emph{all} parameters using a computational effort similar to that of evaluating the forward problem $\mathcal{L}$ \emph{once}. More detail on this calculation can be found in Appendix~\ref{appendix}.

\section{Results and Discussion}
\label{section_results}

\begin{figure*}
    \centering
    \includegraphics[width=0.85\linewidth]{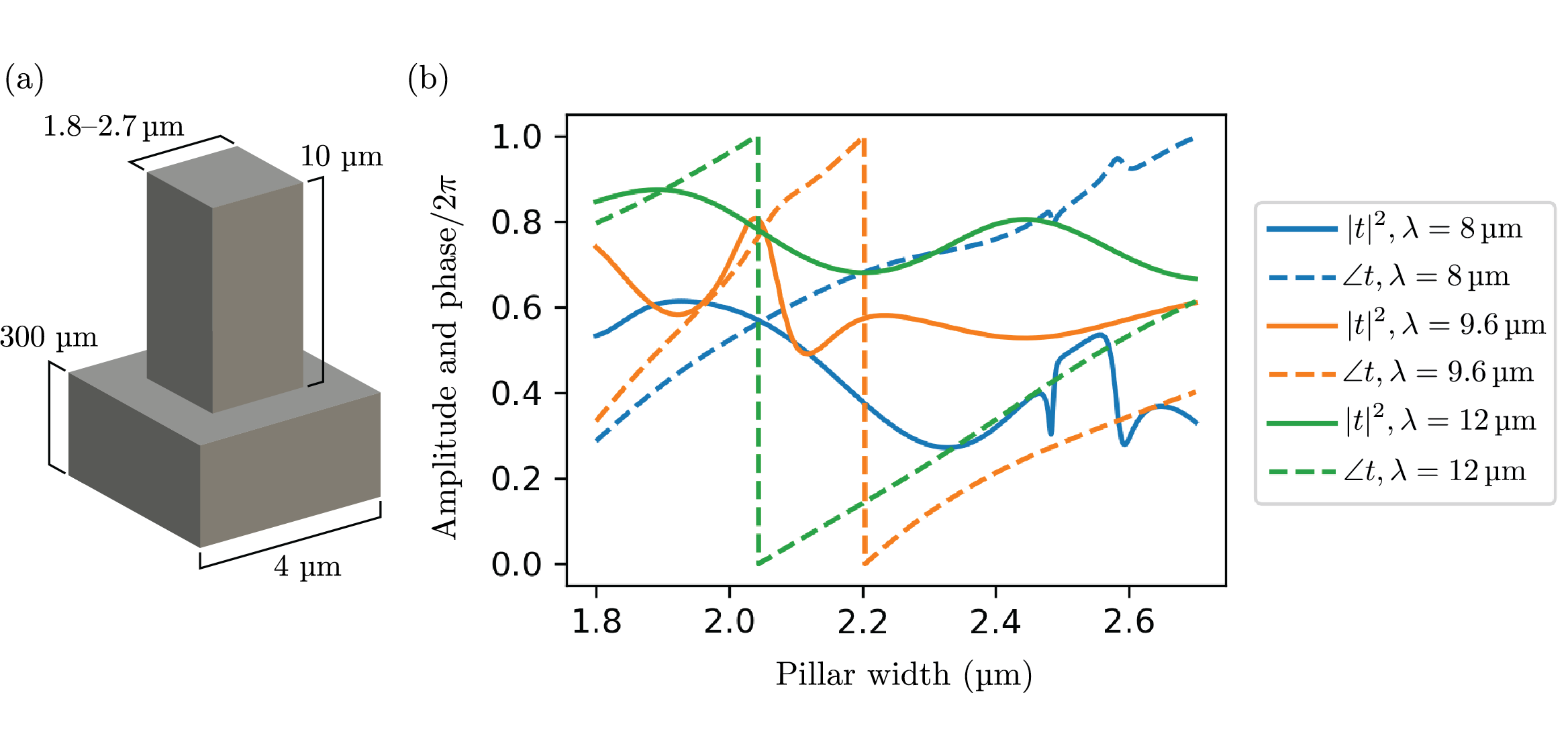}
    \caption{ (a) The metasurface unit cell, consisting of a square silicon pillar with a variable width of \SIrange[range-phrase=\textendash, range-units=single]{1.8}{2.7}{\micro\metre} on a silicon substrate. (b) The phase $\angle t$ and the amplitude squared  $T = |t|^2$ of the unit cell complex transmission coefficient $t$ as a function of pillar width, calculated at the center frequency $(\lambda = \SI{9.6}{\micro\metre})$ and at the  bandwidth limits $(\lambda = \SI{8}{\micro\metre}, \lambda = \SI{12}{\micro\metre})$. }
    \label{results_geometry_figure}
\end{figure*}

\begin{figure*}
    \centering
    \includegraphics[width=\linewidth]{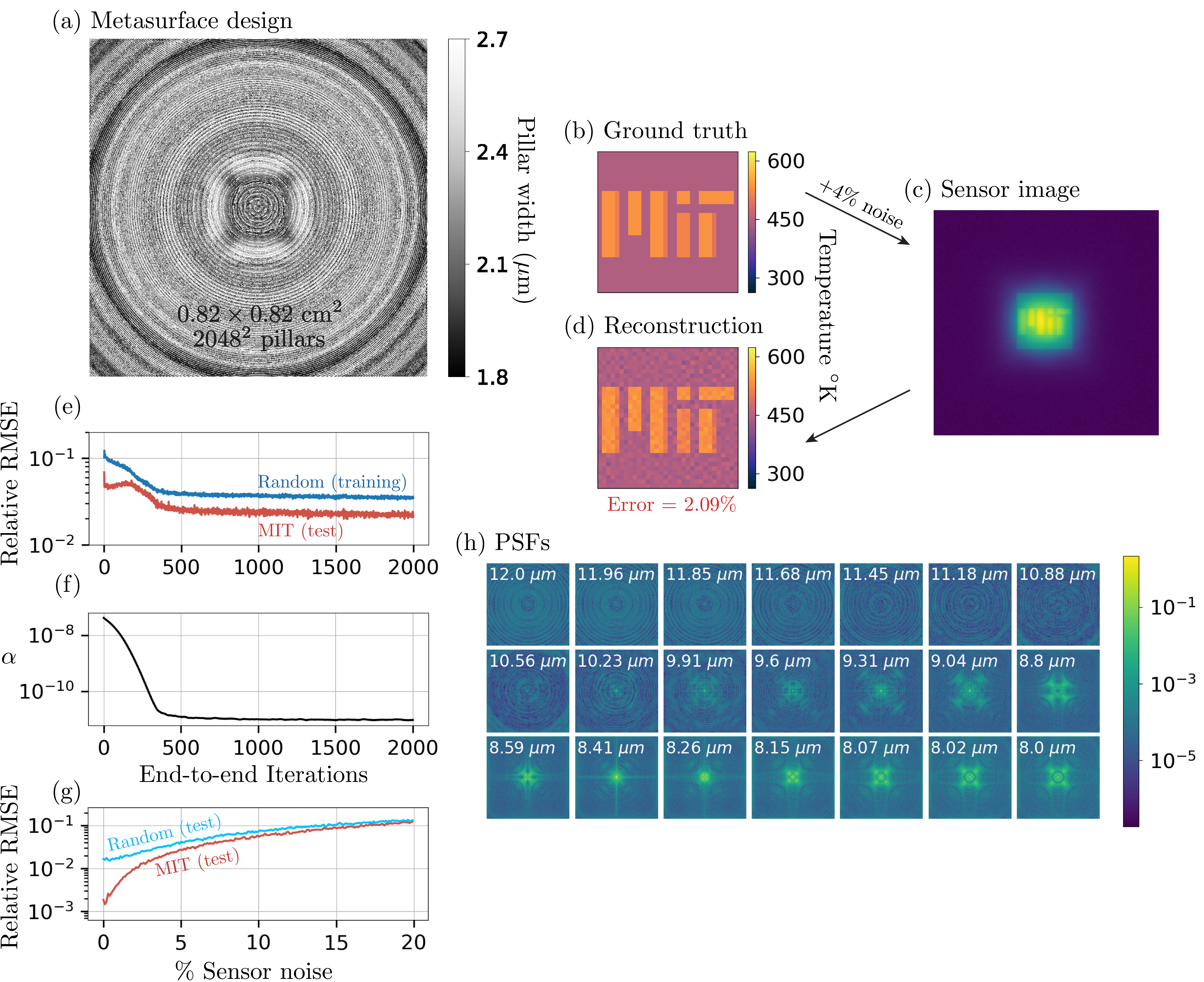}
    \caption{An end-to-end optimized single-layer metasurface design used to reconstruct temperature maps of $32^2$ pixels from $128^2$ monochrome sensor images of LWIR thermal radiation (\SIrange[range-phrase=\textendash, range-units=single]{8}{12}{\micro\metre}), with a sensor pixel size of $\SI{12}{\micro\metre}$. The metasurface consists of $2048^2$ unit cells with a $\SI{4}{\micro\metre}$ period (Figure~\ref{results_geometry_figure}), and the NA is $0.20$. In terms of the wavelength corresponding to the center frequency, $\lambda = \SI{9.6}{\micro\metre}$, the metasurface diameter is $853.33\lambda$, and the distance from the metasurface to the sensor is $2083.33 \lambda$. In the end-to-end optimization, the metasurface is initialized as an inverse-designed monochromatic lens at the center frequency, and converges to the structure depicted in (a), which plots the width of each pillar over the metasurface area. (b--d): The end-to-end design accurately reconstructs temperature maps of arbitrary objects such as the MIT logo, with a relative RMSE (root mean square error) of $2.09 \%$ for $4 \%$ sensor noise. (e): The evolution of the objective function [Equation~\eqref{endtoend_objective}] (reconstruction of random temperature maps, blue) and the reconstruction error of a test MIT logo  (red) throughout the end-to-end optimization, and (f) the evolution of the reconstruction hyper-parameter $\alpha$. (g): Using the end-to-end optimized design, the reconstruction error vs.~sensor noise amplitude for a random test temperature map (light blue) and the test MIT logo (red). (h): The PSFs of the metasurface at the 21 Chebyshev points used for the spectral integration scheme of the image formation~[Equation~\eqref{imageformation}]. }
    \label{results1_figure}
\end{figure*}

\begin{figure*}
    \centering
    \includegraphics[width=0.85\linewidth]{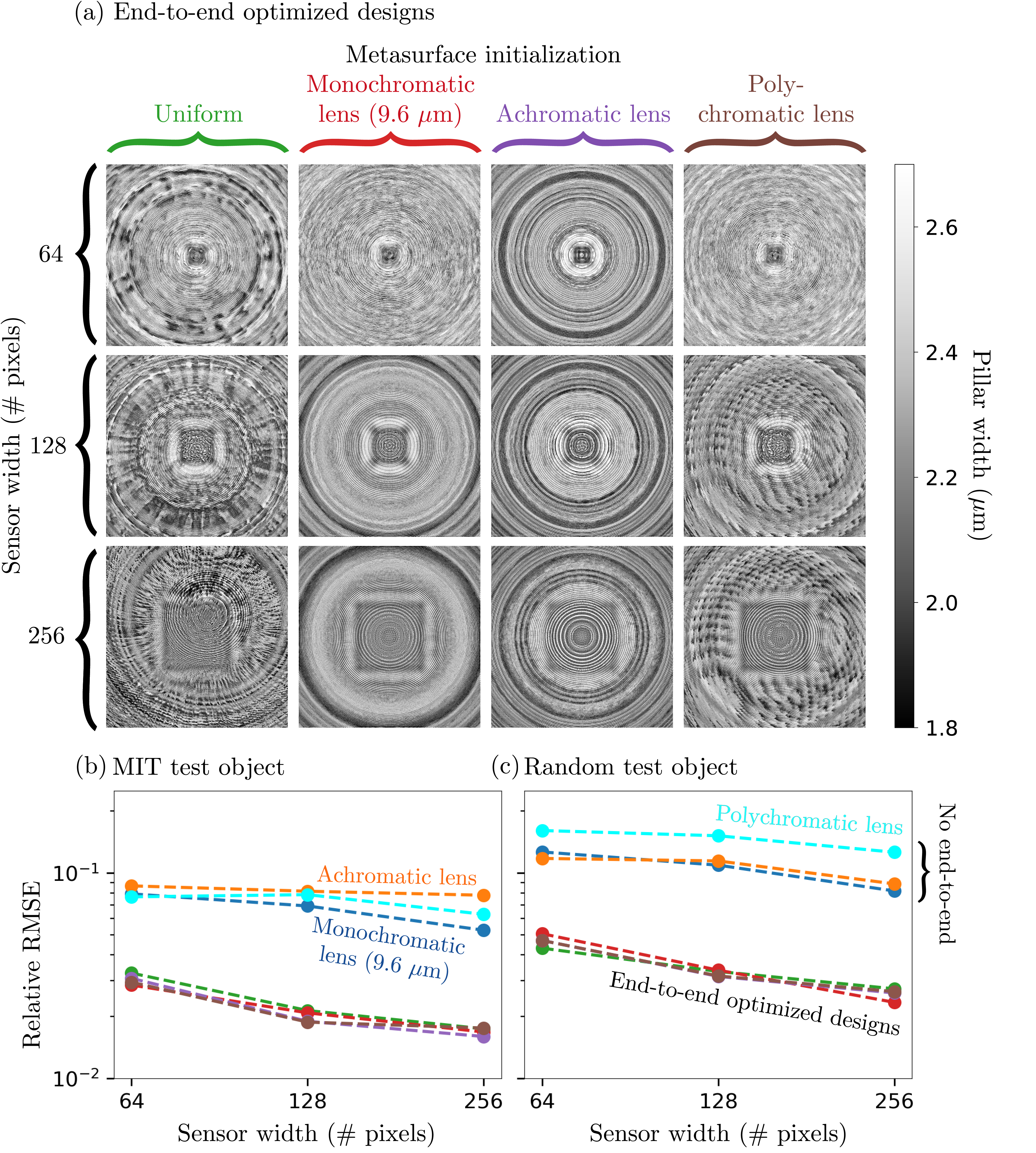}
    \caption{(a): End-to-end optimized single-layer metasurface designs for varying sensor sizes: $64^2$, $128^2$, and $256^2$ sensor pixels, and varying metasurface initalizations: uniform pillar widths, and three inverse-designed metalenses---a monochromatic lens optimized to maximize focal intensity~\cite{pestourieInverseDesignLargearea2018} at the center frequency $\lambda = \SI{9.6}{\micro\metre}$, an ``achromatic'' lens optimized to maximize the worst-case focal intensity~\cite{pestourieInverseDesignLargearea2018,liInverseDesignEnables2022} over the whole LWIR bandwidth, and a polychromatic lens optimized to focus five different frequencies to five different spots on the sensor. The parameters of the metasurfaces are the same as in Figure~\ref{results1_figure}. For the three sensor sizes, we apply $2 \%, 4 \%$ and $8 \%$ noise (relative to the image mean) since we observe that the image mean roughly decreases by half when we double the sensor width. (b and c): The relative RMSE of the reconstructions of a test MIT logo and a test random object for the end-to-end designs shown in (a). For a fixed sensor size, the end-to-end designs resulting from different metasurface initializations all have similar performance and improve upon the inverse-designed baselines (initial inverse-designed metalenses combined with the Planck regression). The reconstruction error of the end-to-end designs decreases as the sensor size increases. }
    \label{results2_figure}
\end{figure*}

We now present the results of our end-to-end optimization framework applied to an example metasurface system, demonstrating high-quality temperature-map reconstruction from a grayscale thermal sensor image through the co-design of a single-layer metasurface and our Planck reconstruction algorithm. We compare our results to various baselines, such as inverse-designed metasurfaces that are designed \textit{independently} of the reconstruction (no end-to-end design), as well as a convolutional neural network (CNN) reconstruction algorithm. Our metasurfaces are composed of a grid of $2048 \times 2048$ silicon unit cells on a silicon substrate.
Figure~\ref{results_geometry_figure}a depicts the unit cell geometry, which has a period of $\SI{4}{\micro\metre}$ and contains a square pillar with a height of 
 $\SI{10}{\micro\metre}$ and a variable width of $\SIrange[range-phrase=\textendash, range-units=single]{1.8}{2.7}{\micro\metre}$; these parameters are based on previous experimental LWIR metasurfaces~\cite{huangLongWavelengthInfrared2021}. The metasurface optimization parameters $\vb{p}$ are therefore a $2048^2$-component vector representing the widths of all pillars. (More generally, one could include multiple degrees of freedom per unit cell~\cite{liEmpoweringMetasurfacesInverse2022}, e.g.~rectangular pillars~\cite{liInverseDesignEnables2022}.) Figure~\ref{results_geometry_figure}b plots the phase and amplitude squared of the unit cell transmission coefficient $t$ as a function of the pillar width for three different wavelengths in the bandwidth, showing that there is significant dispersion (wavelength dependence), and with these limited degrees of freedom a full $[0,2\pi)$ phase coverage is not attained at all $\lambda$. Thus, we expect a poor-quality achromatic lens from such a metasurface, but nevertheless our algorithm recovers an accurate diffraction-limited reconstruction below.
 
 We consider temperatures ranging from $T_{\text{low}} = \SI{263.15}{\kelvin}$ to $T_{\text{high}} = \SI{623.15}{\kelvin}$ (a typical range for conventional thermal cameras), and we set $T_{\text{bg}}$ from Equation~\eqref{reconstruction} to be at the center of the range, $T_{\text{bg}} = \SI{443.15}{\kelvin}$ (in practice, one would probably use a measured ambient temperature). We reconstruct $32 \times 32$ temperature maps and consider a few different problems of varying sensor size in the over-determined regime, where the number of sensor pixels is greater than the number of object pixels (see Figure~\ref{results2_figure}). In all of the problems considered here, the numerical aperture is $\text{NA} = 0.20$ and the sensor pixel size is $\SI{12}{\micro\metre}$, below the range of diffraction limits of the system ($\sim \SIrange[range-phrase=\textendash, range-units=single]{20}{30}{\micro\metre}$).

 Figure~\ref{results1_figure} examines the performance of one particular end-to-end design, where the sensor size is $128^2$ pixels and the metasurface is initialized as an inverse-designed monochromatic lens at the center frequency ($\lambda = \SI{9.6}{\micro\metre})$. The end-to-end optimization~(\ref{endtoend_objective}) converges well within $2000$ iterations (Figure~\ref{results1_figure}e) to a relative RMSE (the square root of the relative MSE $\mathcal{L}$) of $\sim 4 \%$. Figure~\ref{results1_figure}a shows the optimized metasurface, which retains some of the lens-like features of the initial structure but exhibits notable variations in the overall design. The corresponding optimized PSFs are plotted in Figure~\ref{results1_figure}h at a set of 21 Chebyshev points~\cite{trefethenApproximationTheoryApproximation2013} in the spectral bandwidth [used for the numerical integration of~\eqref{imageformation}]. The optimized metasurface retains a near diffraction-limited focal spot at the center frequency, but develops additional focal spots of comparable or higher intensity in the $\SIrange[range-phrase=\textendash, range-units=single]{8}{10.56}{\micro\metre}$ bandwidth. Compared with the monochromatic lens, many of these focal spots exhibit a central lobe with a slightly smaller width in addition to higher-intensity sidelobes. Above $\SI{10.56}{\micro\metre}$, the PSFs fall significantly in intensity and exhibit large spatial dispersion. Figure~\ref{results1_figure}f plots the evolution of the reconstruction hyper-parameter $\alpha$ during the optimization. We see that $\alpha$ decreases from $\sim 2 \times 10^{-8}$ to $\sim 2 \times 10^{-11}$ which is small enough such that the regularization is no longer needed, suggesting that the end-to-end optimization improves the condition number of the reconstruction problem. The initial value of $\alpha$ is obtained by a crude pre-optimization: given the initial inverse-designed lens and a fixed random temperature map, we solve for the value of $\alpha$ that minimizes the initial reconstruction error. This allows us to directly compare the reconstructions of the initial design with the end-to-end optimized design. 
 
 Although our optimized metasurface is trained on random temperature maps, the Planck reconstruction also generalizes to perform well (using the same metasurface) on completely different, non-random temperature maps, such as that shown in Figure~\ref{results1_figure}b (the ``MIT'' logo against a background temperature of $T_{\text{bg}}$, where the logo is between $T_{\text{bg}}$ and $T_{\text{high}}$). Figures~\ref{results1_figure}c and~d show the sensor image formed with $4\%$ noise added and the reconstructed MIT logo with $2.09 \%$ error (relative RMSE). We also examine the scaling of the reconstruction error with the sensor noise for both the MIT logo and a random test object (Figure~\ref{results1_figure}g). Along with the objective function, Figure~\ref{results1_figure}e plots the reconstruction error of the test MIT logo throughout the optimization. The objective function improves from over $\sim 10 \%$ to $\sim 4 \%$, while the reconstruction error of the MIT logo improves from $\sim 7 \%$ to $\sim 2 \%$. Combining an inverse-designed monochromatic lens with the reconstruction algorithm already results in good quality reconstructions with relatively low error. That is, without doing any end-to-end design, the reconstruction problem is already fairly accurate due to the strong physical priors of the blackbody emission and image formation, but the end-to-end optimization is able to further improve the reconstruction error by several times for both the random training objects and test MIT logo.

 % MERGE TO CONCLUSIONS:
 %When examining the performance of the end-to-end design in different parameter regimes, increasing the sensor pixel size to be well over the diffraction limit makes the problem ``too easy'' for Planck regression, so end-to-end design cannot improve over the initial metalens.  Conversely, the fact that Planck regression does so well suggests that one should be able to tackle even more challenging problems in the future, such as reconstruction of depth information; by increasing the difficulty of the problem, we expect that end-to-end design will have an even bigger impact.

Our model of the blackbody physics (Section~\ref{section_theory}) assumes that the objects have emissivities $\varepsilon_{f}(x,y) \approx 1$ (corresponding to materials absorbing nearly all LWIR light), but violating this assumption introduces only modest errors. We quantified this by studing the performance of the Figure~\ref{results1_figure} design in reconstructing objects with emissivities significantly below unity. For instance, with $\varepsilon_{f}(x,y) = 0.8$ (a 20\% error in emissivity), a random temperature map is reconstructed with only $6.73 \%$ error in the temperature, and the ``MIT'' temperature map with $5.98 \%$ error. Moreover, the reconstructions are visually similar to the ground truth objects but with systematic errors in the reconstructed temperatures. We also considered the reconstruction of an MIT object where the $20\%$ emissivity error is only applied to the letters of the logo rather than to both the logo and the background; in this case, the overall temperature error is  $3.84 \%$, in the form of a systematic error in only the temperatures of the logo letters.  Mathematically, these results are not surprising: the emissivity multiplies the blackbody emission~(\ref{planckslaw}) linearly, whereas the temperature enters \emph{exponentially}, so we anticipate that an error in the emissivity should have only a logarithmic effect on the inferred temperature.

 \begin{figure*}
    \centering
    \includegraphics[width=0.7\linewidth]{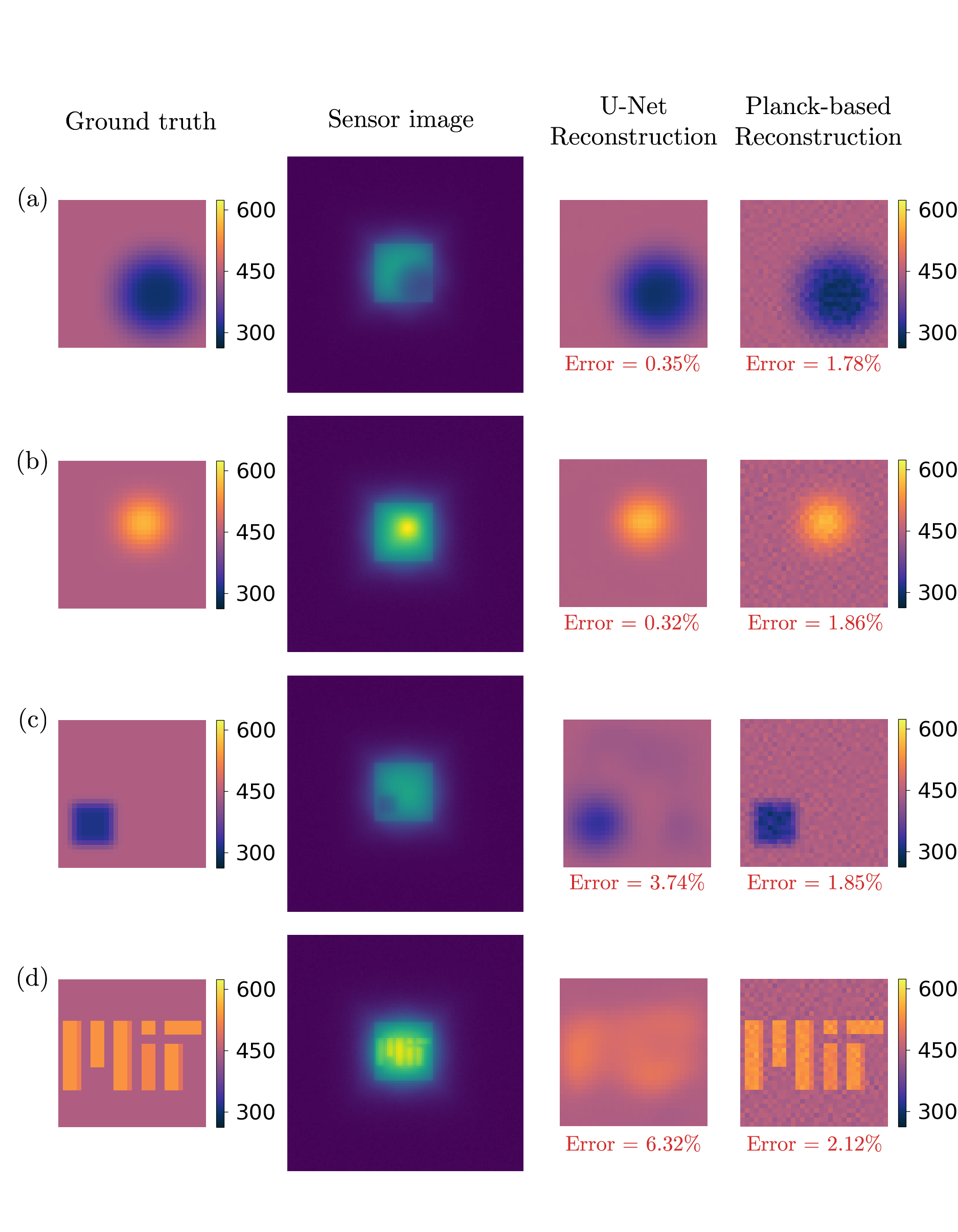}
    \caption{Reconstruction of various test temperature maps using both the Planck-based reconstruction algorithm and a baseline CNN reconstruction algorithm with a U-Net architecture. The sensor images are generated using the end-to-end designed metasurface from  Figure~\ref{results1_figure} with $4 \%$ sensor noise. The CNN is trained on a set of 52,000 images of circles with randomly sampled radii and temperature between $T_{\text{low}} = \SI{263.15}{\kelvin}$ to $T_{\text{high}} = \SI{623.15}{\kelvin}$, over a constant background $T_{\text{bg}} = \SI{443.15}{\kelvin}$. (a) and (b): Reconstructions of ``test” circles that are similar to the CNN training set, for which the U-Net reconstruction is extremely accurate and outperforms the Planck-based algorithm. (c) and (d): Reconstructions of test temperature maps that are different from the CNN training set: a square and the MIT logo against constant backgrounds of $T_{\text{bg}}$. The U-Net fails to generalize to these cases and produces unrecognizable reconstructions, in contrast to the Planck-based algorithm which demonstrates accurate and recognizable reconstructions.}
    \label{NN_figure}
\end{figure*}

In Figure~\ref{results2_figure}, we examine the effects of metasurface initialization and sensor size on the end-to-end performance. We consider four different metasurface initializations: uniform pillar widths, a monochromatic lens optimized to maximize focal intensity~\cite{pestourieInverseDesignLargearea2018} at the center frequency $\lambda = \SI{9.6}{\micro\metre}$, an ``achromatic'' lens optimized to maximize the worst-case focal intensity~\cite{pestourieInverseDesignLargearea2018,liInverseDesignEnables2022} over the whole LWIR bandwidth, and a polychromatic lens optimized to focus five different frequencies to five different spots on the sensor (again maximizing the worst-case intensity at each spot) inspired by hyperspectral metasurface designs~\cite{linEndtoendMetasurfaceInverse2022}. We consider three different sensor sizes: ${64}^2$, ${128}^2$, and ${256}^2$ sensor pixels. For sensors with $64^2$ pixels, we apply $2\%$ noise to the images (applied relative to the image mean); as we double the sensor width, we also double the sensor noise since we observe that the image mean roughly decreases by half (because most of the PSF pixels are dark). Since the end-to-end optimization is very non-convex, we observe that different metasurface initializations lead to very different designs (Figure~\ref{results2_figure}a), although many have qualitatively similar focusing properties to the example described in Figure~\ref{results1_figure}. For a fixed sensor size, however, the different designs nevertheless achieve very similar reconstruction accuracy (Figures~\ref{results2_figure}b and~c). Figures~\ref{results2_figure}b and~c examine the scaling of reconstruction errors with sensor sizes for both the inverse-designed and end-to-end metasurfaces. From $64^2$ to $128^2$ sensor pixels, the end-to-end reconstruction error decreases relative to the initial reconstruction error, but there is less relative improvement from $128^2$ to $256^2$ (i.e.~the errors of both the initial and end-to-end designs decrease by about the same amount). Even at larger sensor sizes, the end-to-end designs still make focal spots that lie within a small portion of the sensor, which is perhaps why we do not consistently see the reconstruction error improve as the sensor area is increased.

\subsection{CNN baseline comparison}

Next, as a baseline comparison to our framework, we also consider a CNN reconstruction algorithm~\cite{mccannConvolutionalNeuralNetworks2017}, which is essentially a form of ``black-box'' nonlinear function (containing no physics knowledge) that is trained to reconstruct the ground-truth images of a training set.   There are innumerable variations of neural-network (NN) architectures, but a convolutional NN (CNN) is a common choice for image processing because it captures the translation invariance of image problems (i.e.~if the input image is shifted, the output reconstruction should also be shifted).  As described in detail below, we find that our baseline CNN performs poorly: it cannot reconstruct random images at all, and succeeds only on a highly constrained training set consisting of images of circles.   While it is conceivable that some other NN architecture could perform better, or that one could succeed with a vastly larger NN coupled with an enormous training set, our baseline CNN results underscore the advantage that is gained by incorporating physical knowledge of thermal emission and scattering (PSFs) into the reconstruction algorithm. 

In particular, we consider the combination of our end-to-end designed metasurface (see above) coupled with a CNN reconstruction, and we train the CNN to reconstruct temperature maps given sensor images formed by the metasurface. (We emphasize that no end-to-end optimization is performed for the combined metasurface and CNN reconstruction. Recall that Planck reconstruction performed well even without end-to-end optimization, so we fix the metasurface to be the output of our previous end-to-end design, replace the Planck regression with a CNN and optimize the CNN parameters only.) In particular, we consider the metasurface from Figure~\ref{results1_figure}, which is designed for reconstructing temperature maps with $32^2$ pixels from sensor images with $128^2$ pixels and $4\%$ sensor noise. We employ a CNN based on the U-Net architecture with a standard encoder--decoder structure~\cite{ronnebergerUNetConvolutionalNetworks2015}.   

Above, Planck regression (with or without end-to-end meta-optics optimization) successfully reconstructed temperature maps sampled from uniform random pixels---that is, assuming no structure besides bounds on the allowed temperatures.  We tried training the CNN on the same type of data, with 52,000 random images, but we found that the CNN converged to a result that had nearly 100\% reconstruction error, i.e.~training was not able to improve its loss function.  Typically NNs are applied to much more constrained training sets, e.g.~based on sets of real-world images, which makes the problem effectively lower dimensional.  To simplify the problem into something tractable for the CNN, therefore, we tried training it on a much more structured dataset of the same size, consisting of circles with randomly sampled radii and temperatures (from $T_{\text{low}}$ to $T_{\text{high}}$) over a constant background $T_{\text{bg}}$. In this simplified case, the CNN successfully learned from the data, but its reconstruction did not generalize to other types of images, as explained in detail below.

Figure~\ref{NN_figure} shows the performance of the trained CNN compared with our Planck regression on various test temperature images. When we evaluate the CNN on ``circle'' temperature images that are similar to the training set, the CNN is able to reconstruct them with very high accuracy, even with 4\% sensor noise. Figure~\ref{NN_figure} shows two such examples, where the CNN has reconstruction errors of $0.35\%$ and $0.32\%$ (relative RMSE). The Planck regression also shows high accuracy on these examples with errors of $1.78\%$ and $1.86\%$, although the errors are greater than that of the NN because the end-to-end was not specialized to this type of image. However, when we evaluate the two algorithms on test data that is unlike the training set, the CNN has poor performance while the Planck regression performs well. For instance, Figure~\ref{NN_figure} shows the reconstruction of a square against a flat background as well as the MIT logo; the CNN has reconstruction errors of $3.74\%$ and $6.32\%$ respectively, while the Planck regression has errors of $1.85\%$ and $2.12\%$.  Moreover, the CNN reconstructions are blurry and nearly unrecognizable, while Planck regression produces clearly recognizable images in which the errors take the appearance of white noise (that could be reduced by additional de-noising algorithms~\cite{fanBriefReviewImage2019}). 
While our optimization of the CNN was not exhaustive, our results are generally consistent with the literature on CNN image reconstruction, where it is only for \emph{in-distribution} reconstruction (where the test images are within the low-dimensional distribution of images used in the training set) that image recovery with very low error is possible~\cite{xuDeepConvolutionalNeural2014, mccannConvolutionalNeuralNetworks2017}. On the other hand, for out-of-distribution (OOD) reconstruction (in which the test images are significantly different from those used in the training set), a much larger reconstruction error is typical~\cite{osadaOutDistributionDetectionReconstruction2023a}. Moreover, no CNN has demonstrated reconstruction with perfect image recovery for completely random (white noise) images. In contrast, physics-based reconstruction does not require training and generalizes to arbitrary images.

\section{Conclusion}
In summary, we have presented a framework for temperature map sensing from LWIR thermal radiation through the co-optimization of metasurface optics and a nonlinear Planck-regression \eqref{planckslaw} algorithm. Our framework demonstrates high-quality reconstructions of any temperature map, including completely random ones. The example metasurfaces designed in this work are experimentally feasible~\cite{fanHighNumericalAperture2018, huangLongWavelengthInfrared2021, kignerMonolithicAllsiliconFlat2021, linWideFieldofViewLargeAreaLongWave2024, meemBroadbandLightweightFlat2019, wirth-singhLargeFieldofviewThermal2023}, composed of square silicon pillars that display high transmission in the LWIR bandwidth and are fabricable with standard techniques, but are only one of many possible optical geometries that could be used with our methods. For example, one could conceivably achieve even higher performance by increasing the complexity of the metasurface geometry to allow additional degrees of freedom per unit cell~\cite{liAdvancesExploitingDegrees2020} or by employing more sophisticated optical modeling that goes beyond the locally periodic approximation~\cite{linOverlappingDomainsTopology2019, christiansenFullwaveMaxwellInverse2020}. 

Our approach overcomes a major challenge of conventional thermal-imaging systems, which require bulky optics to achieve high-quality, achromatic imaging. Metasurfaces provide a promising platform to miniaturize these systems, but single-layer metasurfaces are limited in their broadband performance by fundamental delay-bandwidth constraints~\cite{presuttiFocusingBandwidthAchromatic2020}. Multi-layer metasurfaces (volumetric meta-optics) offer additional degrees of freedom that can realize high-quality achromatic optics but introduce complicated three-dimensional nano-fabrication challenges~\cite{linComputationalInverseDesign2021}. By synergizing the metasurface design with computation, the end-to-end approach \emph{bypasses} such physical limitations, enabling high-quality temperature map reconstruction with only a single layer. We emphasize that, in the end-to-end approach, the metasurface on its own does not produce a usable image; rather it is the \emph{combination} of the metasurface and the Planck regression that yield a meaningful reconstruction. Designing the optics and the image processing together allows for maximal exploitation of the blackbody physics ``prior'' that is unique to the thermal regime. 

In modeling the image formation, we assumed an additive Gaussian white noise distribution for the sensor noise, whereas thermal sensors typically have non-uniform additive and multiplicative noises due to internal heating and stray reflections of the camera, as well as additive readout noise. One extension of this work could be to incorporate such noise models into the image formation. As previous works have considered, the non-uniform multiplicative and additive noise terms may even be reconstructed along with the temperature map, although this typically requires multiple shots of the same scene (which harnesses the fact that these terms change slowly over time for a given camera) \cite{saragadamThermalImageProcessing2021, hardieMAPEstimatorSimultaneous2007, hardieScenebasedNonuniformityCorrection2000}. One may also consider more realistic properties of a thermal sensor affecting image formation such as spectral sensitivity of the sensor. 

As discussed in Section~\ref{section_results}, the systems considered in this work have a sensor pixel size that is below the range of diffraction limits of the system at these wavelengths, and the reconstructed objects thus have the highest resolution possible.  (The problem becomes inherently ill-conditioned if one attempts to further increase the resolution into the sub-diffraction regime, and such super-resolution imaging would require additional priors~\cite{tianSurveySuperresolutionImaging2011}.) Moreover, for such systems, we found that even before doing any end-to-end design of the metasurface (but instead using simple inverse-designed metalenses), the reconstruction is already fairly accurate due to its strong physical foundations. We found that when we increase the sensor pixel size to be larger than the diffraction limit, thereby decreasing the object resolution, the reconstruction problem becomes ``too easy'' and the end-to-end design does not improve upon the initialized system. In general, other possible extensions of this work include exploring regimes where the reconstruction problem is even more challenging and where we expect end-to-end design to have an even greater impact. One example of this is the reconstruction of larger temperature maps in higher-NA regimes. Since LPA empirically breaks down beyond a certain NA~\cite{pestourieInverseDesignLargearea2018, perez-arancibiaSidewaysAdiabaticityRay2018, liInverseDesignEnables2022}, this may require more accurate modeling that captures additional wave physics~\cite{linOverlappingDomainsTopology2019, christiansenFullwaveMaxwellInverse2020}. Moreover, since the problems considered here are overdetermined, increasing the temperature-map size significantly may also require increasing the metasurface area. In cases where the PSFs are localized towards the center of the sensor (as is the case for the metasurfaces in Figures~\ref{results1_figure} and \ref{results2_figure}), it may be possible to increase the object resolution without re-designing the metasurface. For the metasurface in Figure~\ref{results1_figure}, we found that when we double the size of the object from $32^2$ to $64^2$ total pixels, keeping the metasurface geometry and the sensor size fixed, the reconstruction error only increases by $\sim 1 \%$.  While increases in object size would require a larger metasurface and sensor, we expect similar accuracy from the reconstruction, because our model is translation-invariant and the PSFs of even a poor lens are localized (so that the reconstruction accuracy at widely separated pixels is decoupled).  The computational cost scales roughly proportional to the area. 

Alternatively, one could extend this framework to underdetermined regimes (reconstructing images larger than the sensor) by applying additional priors such as sparsity~\cite{aryaEndtoEndOptimizationMetasurfaces2024}. Another way to make the problem more difficult would be to consider 3D temperature map reconstruction, e.g.~simultaneous reconstruction of temperature [$T(x,y)$] and depth [$z(x,y)$] maps. In Section~\ref{section_results}, we found that errors in the emissivity introduce only small errors in the reconstructed temperature.  One could conceivably increase the accuracy of the model as well as the difficulty of the problem by attempting to reconstruct the (wavelength-dependent) emissivity of an object in addition to its temperature map.   

Although we found that a typical convolutional neural network (CNN) architecture could not be trained to reconstruct general images, it is possible that more sophisticated machine-learning approaches could be devised, perhaps by incorporating physical models directly into the NN or by using a CNN as an additional regularization term in a physics-based reconstruction~\cite{saragadamFoveatedThermalComputational2024, saragadamThermalImageProcessing2021, huangBroadbandThermalImaging2024}.  The potential advantage of a NN in this context is that it could be used to express additional knowledge about the set of possible images to be reconstructed~\cite{ulyanovDeepImagePrior2020} that can only be formulated in terms of a training set of natural images.

%\medskip
%\textbf{Supporting Information} \par %Please delete the Suppporting Information statement if it is not applicable. Please supply Supporting Information in another file. Supporting information should not be provided in .tex format
%Supporting Information is available from the Wiley Online Library or from the author.

% Acknowledgements
\medskip
\subsection*{Acknowledgements} \par %delete if not applicable))
This work was supported in part by the Simons Foundation,
by the U.S. Army Research Office (ARO) through the Institute for Soldier Nanotechnologies (ISN) under award no.~W911NF-18-2-0048, by the U.S. Army DEVCOM Army Research Laboratory ARO through the MIT ISN under cooperative agreement no.~W911NF-23-2-0121.  AM was supported in part by grant no.~2127331 from the National Science Foundation. ZL was supported in part by grant no.~DE-SC0024223 from the Department of Energy. We are also grateful to Youssef Mroueh at the MIT--IBM Watson AI Lab for advice on CNN reconstruction.

\medskip
\section*{Appendices}

\appendix

\section{Gradient Calculations\label{appendix}}
Here we analytically calculate the end-to-end optimization gradients of Equation~\eqref{endtoend_objective}, denoted by $\partial \mathcal{L} / \partial \vb{p}$ and $\partial \mathcal{L} / \partial \alpha$. We can omit the averages over the training set and noise profiles in the objective function, because stochastic gradient-descent algorithms require only the gradients for a single sample~\cite{kingmaAdamMethodStochastic2014} (temperature map $\vb{T}$ and noise profile $\vb*{\eta}$). We let $N^2$ be the dimension of the temperature maps $\vb{T}$ and $\vb{T}_{\text{est}}$, $M^2$ be the dimension of the sensor image $\vb{v}$, $P^2$ be the dimension of the metasurface parameter vector $\vb{p}$, $R^2$ be the number of metasurface unit cells, and $B^2$ be the number of adjacent image pixels that are binned together to form the sensor image (in the problems considered here, $N = 32, M = 64-256, P = 2048, R = 2048$, and $B = 3$). We begin by applying the chain rule to the gradients:
\begin{align}
    \frac{\partial \mathcal{L}}{\partial \vb{p}} &=  \frac{\partial \mathcal{L}}{\partial \vb{T}_{\text{est}}}  \frac{\partial \vb{T}_{\text{est}}}{\partial \vb{p}}  \label{chainrule1}\\ 
    \frac{\partial \mathcal{L}}{\partial \alpha} &=  \frac{\partial \mathcal{L}}{\partial \vb{T}_{\text{est}}}  \frac{\partial \vb{T}_{\text{est}}}{\partial \alpha} \label{chainrule2}.
\end{align}
$\partial \mathcal{L} / \partial \vb{T}_{\text{est}}$ is found by directly differentiating Equation~\eqref{endtoend_objective}. To compute $\partial \vb{T}_{\text{est}} / \partial \vb{p}$ and $\partial \vb{T}_{\text{est}} / \partial \alpha$, we apply the KKT optimality conditions~\cite{boydConvexOptimization2004} to the reconstruction [Equation~\eqref{reconstruction}], which in this case are simply the condition that the gradient of the objective function with respect to $\vb*{\tau}$ is zero at the optimum ($\vb*{\tau} = \vb{T}_{\text{est}}$):
\begin{equation}
    2 \alpha \vb{T}_{\text{est}} - 2 \bigg( \int_{\Delta f} df \, \frac{\partial \vb{b}(f, \vb*{\tau})}{\partial \vb*{\tau}} 
    \bigg|_{\vb{T}_{\text{est}}}  \hat{G}(f, \vb{p} )^{\mathsf{T}} \bigg) \bigg( \vb{v} -  \int_{\Delta f} df \, \hat{G}(f, \vb{p} ) \vb{b}(f,  \vb{T}_{\text{est}})  \bigg) = \vb{g}(\vb{T}_{\text{est}}, \vb{p}) = 0
    \label{KKT}
\end{equation}
where we have labeled the expression on the left hand side as $\vb{g}(\vb{T}_{\text{est}}, \vb{p})$, where $\frac{\partial \vb{b}(f, \vb*{\tau})}{\partial \vb*{\tau}} \bigg|_{\vb{T}_{\text{est}}}$ is an $N^2 \times N^2$ diagonal matrix, and where $\hat{G}(f, \vb{p} )^{\mathsf{T}}$ is the transposed convolution operator, which like $\hat{G}(f, \vb{p} )$ can be applied matrix-free through FFT operations (Section~\ref{section_methods}). This expression also assumes that the $\vb*{\tau} \geq 0$ constraints are not tight at the optimum; empirically we observe this to be the case unless the ground truth temperature map has values that are close to 0. Implicitly differentiating~\cite{blondelEfficientModularImplicit2021} Equation~\eqref{KKT} yields expressions for $\partial \vb{T}_{\text{est}} / \partial \vb{p}$ and $\partial \vb{T}_{\text{est}} / \partial \alpha$ in terms of $\vb{g}(\vb{T}_{\text{est}}, \vb{p})$, which we substitute into Equations \eqref{chainrule1} and \eqref{chainrule2}:
\begin{align}
    \frac{\partial \mathcal{L}}{\partial \vb{p}} &=  
    \frac{2 \big( \vb{T} - \vb{T}_{\text{est}}  \big)^{\mathsf{T}} }{ \big\Vert \vb{T} \big\Vert_2^2  }
    \bigg( \frac{\partial \vb{g}}{\partial \vb{T}_{\text{est}}} \bigg)^{-1} \frac{\partial \vb{g}}{\partial \vb{p}} \\ 
    \frac{\partial \mathcal{L}}{\partial \alpha} &=   \frac{2 \big( \vb{T} - \vb{T}_{\text{est}}  \big)^{\mathsf{T}} }{ \big\Vert \vb{T} \big\Vert_2^2  }
    \bigg( \frac{\partial \vb{g}}{\partial \vb{T}_{\text{est}}} \bigg)^{-1} \frac{\partial \vb{g}}{\partial \alpha}.
\end{align}
where $\partial \vb{g} / \partial \vb{T}_{\text{est}}$ is the $N^2 \times N^2$ Hessian in $\vb*{\tau}$ of the reconstruction objective [Equation~\eqref{reconstruction}] evaluated at $\vb{T}_{\text{est}}$, and $\partial \vb{g} / \partial \vb{p}$ is the $N^2 \times P^2$ Jacobian in $\vb*{\tau}$ and $\vb{p}$ evaluated at $\vb{T}_{\text{est}}$. To compute these expressions efficiently, first we apply the adjoint method \cite{strangComputationalScienceEngineering2012}, where we define the adjoint variable $\Lambda$, an $N^2$ -dimensional vector, by
\begin{align}
    \frac{\partial \mathcal{L}}{\partial \vb{p}} &=  \Lambda^{\mathsf{T}} \frac{\partial \vb{g}}{\partial \vb{p}} \label{adjoint1} \\ 
    \frac{\partial \mathcal{L}}{\partial \alpha} &= \Lambda^{\mathsf{T}} \frac{\partial \vb{g}}{\partial \alpha}.
\end{align}
The vector $\Lambda$ thus satisfies:  
\begin{equation}
    \frac{\partial \vb{g}}{ \partial \vb{T}_{\text{est}} } \Lambda = \frac{2 \big( \vb{T} - \vb{T}_{\text{est}}  \big)^{\mathsf{T}} }{ \big\Vert \vb{T} \big\Vert_2^2  },
    \label{adjointmethod}
\end{equation}
where the Hessian $\partial \vb{g} / \partial \vb{T}_{\text{est}}$ is found by differentiating Equation~\eqref{KKT}:
\begin{equation}
\begin{aligned}
    \frac{\partial \vb{g}}{ \partial \vb{T}_{\text{est}} } = 
    2\alpha I &- 2 \bigg( \int_{\Delta f} df \, \frac{\partial^2 \vb{b}(f, \vb*{\tau})}{\partial \vb*{\tau}^2} \bigg|_{\vb{T}_{\text{est}}}  \hat{G}(f,\vb{p})^{\mathsf{T}}  \bigg) \bigg( \vb{v} - \int_{\Delta f} df \, \hat{G}(f, \vb{p}) \vb{b}(f, \vb{T}_{\text{test}}) \bigg) \\
    &+ 2 \bigg( \int_{\Delta f} df \, \frac{\partial \vb{b}(f, \vb*{\tau})}{\partial \vb*{\tau}} 
    \bigg|_{\vb{T}_{\text{est}}} \hat{G}(f,\vb{p})^{\mathsf{T}} \bigg) \bigg( \int_{\Delta f} df \, \hat{G}(f,\vb{p}) \frac{\partial \vb{b}(f, \vb*{\tau})}{\partial \vb*{\tau}} 
    \bigg|_{\vb{T}_{\text{est}}} \bigg).
\end{aligned}
\end{equation}
We solve Equation~\eqref{adjointmethod} by the iterative conjugate-gradient method \cite{strangComputationalScienceEngineering2012} without forming an explicit matrix for $\partial \vb{g} / \partial \vb{T}_{\text{est}}$ (i.e.~using only its action as a linear operator on vectors, via FFTs), which typically converges within $\sim 150$ iterations. 

The next step is to compute an expression for the right hand side of Equation~\eqref{adjoint1}. Since the Jacobian $\partial \vb{g} / \partial \vb{p}$ is quite large $(N^2 \times P^2)$, we avoid constructing it explicitly and instead compute only the vector--Jacobian product (vJp) $\Lambda^{\mathsf{T}} \frac{\partial \vb{g}}{\partial \vb{p}}$. Expanding the vJp by differentiating Equation~\eqref{KKT}, we find
\begin{equation}
\begin{aligned}
    \Lambda^{\mathsf{T}} \frac{\partial \vb{g}}{\partial \vb{p}} = &- 2 \bigg( \int_{\Delta f} df \, \Lambda^{\mathsf{T}} \frac{\partial \vb{b}(f, \vb*{\tau})}{\partial \vb*{\tau}} 
    \bigg|_{\vb{T}_{\text{est}}} \frac{ \partial \hat{G}(f, \vb{p})^{\mathsf{T}}}{\partial \vb{p}} \bigg) \bigg( \vb{v} - \int_{\Delta f} df \, \hat{G}(f, \vb{p}) \vb{b}(f, \vb{T}_{\text{test}}) \bigg) \\ &-2 \bigg( \int_{\Delta f} df \, \Lambda^{\mathsf{T}} \frac{\partial \vb{b}(f, \vb*{\tau})}{\partial \vb*{\tau}} 
    \bigg|_{\vb{T}_{\text{est}}}  \hat{G}(f, \vb{p})^{\mathsf{T}}  \bigg) \bigg( \frac{\partial \vb*{\eta}}{\partial \vb{p}} +  \int_{\Delta f} df \, \frac{ \partial \hat{G}(f, \vb{p})}{\partial \vb{p}} \big[ \vb{b}(f, \vb{T}) - \vb{b}(f, \vb{T}_{\text{est}}) \big] \bigg). \label{VJP}
\end{aligned}
\end{equation}
where $\vb*{\eta}$ depends on $\vb{p}$ since we scale the noise profile by the mean value of the raw image. In particular, we take 
\begin{equation}
    \vb*{\eta} = \beta * \text{mean} \bigg\{ \int_{\Delta f} df \, \hat{G} (f, \vb{p} )   \vb{b}(f, \vb{T})  \bigg\} \vb{N}(0,1)
\end{equation}
where $\beta$ is a small percentage (in the problems considered here, $\beta = 2-8 \%$) and $\vb{N}(0, 1)$ is Gaussian white noise with a mean of $0$ and a standard deviation of $1$. Expanding the derivative of the noise term, Equation~\eqref{VJP} becomes
\begin{equation}
\begin{aligned}
    \Lambda^{\mathsf{T}} \frac{\partial \vb{g}}{\partial \vb{p}} = &- 2 \bigg( \int_{\Delta f} df \, \Lambda^{\mathsf{T}} \frac{\partial \vb{b}(f, \vb*{\tau})}{\partial \vb*{\tau}} 
    \bigg|_{\vb{T}_{\text{est}}} \frac{ \partial \hat{G}(f, \vb{p})^{\mathsf{T}}}{\partial \vb{p}} \bigg) \bigg( \vb{v} - \int_{\Delta f} df \, \hat{G}(f, \vb{p}) \vb{b}(f, \vb{T}_{\text{test}}) \bigg) \\ &-2 \bigg( \int_{\Delta f} df \, \Lambda^{\mathsf{T}} \frac{\partial \vb{b}(f, \vb*{\tau})}{\partial \vb*{\tau}} 
    \bigg|_{\vb{T}_{\text{est}}}  \hat{G}(f, \vb{p})^{\mathsf{T}}  \bigg) \bigg(  \int_{\Delta f} df \, \frac{ \partial \hat{G}(f, \vb{p})}{\partial \vb{p}} \big[ \vb{b}(f, \vb{T}) - \vb{b}(f, \vb{T}_{\text{est}}) \big] \bigg) \\&-\frac{2 \beta}{M^2} \bigg( \int_{\Delta f} df \, \Lambda^{\mathsf{T}} \frac{\partial \vb{b}(f, \vb*{\tau})}{\partial \vb*{\tau}} 
    \bigg|_{\vb{T}_{\text{est}}}  \hat{G}(f, \vb{p})^{\mathsf{T}} \bigg) \vb{N}(0,1) \vb{1}^{\mathsf{T}} \bigg(  \int_{\Delta f} df \, \frac{ \partial \hat{G}(f, \vb{p})}{\partial \vb{p}}  \vb{b}(f, \vb{T}) \bigg), \label{VJP_expanded}
\end{aligned}
\end{equation}
where $\vb{1}$ is an $M^2$-dimensional vector whose entries are all $1$. Computing this expression requires further expansion of $\partial \hat{G}(f, \vb{p}) / \partial \vb{p}$ and $\partial \hat{G}(f, \vb{p})^{\mathsf{T}} / \partial \vb{p}$, derivatives of the convolution and transposed convolution operators. Below, we will specify how to compute terms of the form $\vb{x}^{\mathsf{T}} \frac{ \partial \hat{G}(f, \vb{p})^{\mathsf{T}}}{\partial \vb{p}} \vb{y} = \frac{\partial}{\partial \vb{p}} (\vb{x}^{\mathsf{T}} \hat{G}(f, \vb{p})^{\mathsf{T}} \vb{y})$, where $\vb{x}$ and $\vb{y}$ are $N^2$ and $M^2$-dimensional vectors, and the derivative is applied only to $\hat{G}^{\mathsf{T}}$. This equation can then be applied directly to the first term of Equation~\eqref{VJP_expanded}, and to the second and third terms by taking transposes. For simplicity, we work with differentials rather than explicitly taking derivatives, calculating how an infinitesimal change in $\vb{x}^{\mathsf{T}} \hat{G}(g,\vb{p})^{\mathsf{T}} \vb{y}$ is related to an infinitesimal change in $\vb{p}$. Expanding $\hat{G}^{\mathsf{T}}$ in terms of FFT operations, we have:
\begin{equation}
    \partial \big( \vb{x}^{\mathsf{T}} \hat{G}(f, \vb{p})^{\mathsf{T}} \vb{y} \big) =  \partial \big( \vb{x}^{\mathsf{T}} \text{CROP}_{\text{TL},N+M,N} \mathcal{F} \big\{ \mathcal{F}^{-1} \big\{\text{PAD}_{\text{BR},M,N+M}\vb{y} \big\} \odot \mathcal{F} \big\{ \vb{PSF}(f, \vb{p}) \big\} \big\}  \big),
    \label{differentialG}
\end{equation}
where $\vb{PSF}(f, \vb{p})$ is an $(N+M)^2$-dimensional vector given by Equation~\eqref{PSFs}, $\mathcal{F}$ and $\mathcal{F}^{-1}$ are the 2D Discrete Fourier Transform (DFT) and its inverse; $\odot$ takes the elementwise product of two vectors; $\text{PAD}_{\text{BR}}$ is an operator that zero-pads an $M \times M$ grid to a size of $(N + M) \times (N+M)$ by placing the grid in the bottom right corner, and $\text{CROP}_{\text{TL},N+M,N}$ is an operator that crops an $(N + M) \times (N+M)$ grid to a size of $N \times N$ from the top left corner. We note that since all of the vectors represent flattened 2D grids, the actions of $\mathcal{F}, \mathcal{F}^{-1}, \text{PAD}_{\text{BR},M,N+M}$ and $\text{CROP}_{\text{TL},N+M,N}$ can be described as reshaping the vector into a grid, performing the respective operation on the grid, and then flattening it back into a vector. 

The right hand side of Equation~\eqref{differentialG} depends on $\vb{p}$ only through $\vb{PSF}(f, \vb{p})$, allowing us to propagate the differential to this term. By carrying out the algebra, we can express the right hand side as the dot product of $\partial \vb{PSF}(f, \vb{p})$ with another vector. We find 
\begin{equation}
    \partial \big( \vb{x}^{\mathsf{T}} \hat{G}(f, \vb{p})^{\mathsf{T}} \vb{y} \big) = ( \mathcal{F} \{ \text{PAD}_{\text{TL},N,N+M} \vb{x} \} \odot  \mathcal{F}^{-1} \{ \text{PAD}_{\text{BR},M,N+M}\vb{y} \})^{\mathsf{T}} \partial \vb{PSF}(f, \vb{p}) 
    \label{differentialG2}
\end{equation}
where we have used the fact that $\mathcal{F}^{\mathsf{T}} = \mathcal{F}$, and that the transpose of a crop operation is a zero-pad (and vice versa): $\text{CROP}_{\text{TL},N+M,N}^{\mathsf{T}} = \text{PAD}_{\text{TL},N,N+M}$, where $\text{PAD}_{\text{TL},N,N+M}$ is the operator that zero-pads an $N \times N$ grid to a size of $(N + M) \times (N + M)$ by placing the grid in the top left corner. 

The next steps are to expand $\partial \vb{PSF}(f, \vb{p})$ according to Equation~\eqref{PSFs} and express the near-to-far-field transformation $\text{FF}(\cdot)$ in terms of FFT operations. Equation~\eqref{differentialG2} becomes:
\begin{equation}
\begin{aligned}
    \partial \big( \vb{x}^{\mathsf{T}} \hat{G}&(f, \vb{p})^{\mathsf{T}} \vb{y} \big) = ( \mathcal{F} \{ \text{PAD}_{\text{TL},N,N+M} \vb{x} \} \odot  \mathcal{F}^{-1} \{ \text{PAD}_{\text{BR},M,N+M}\vb{y} \} )^{\mathsf{T}} \cdot \\ &\partial \bigg( \frac{1}{f} \text{BIN}_B  \big| \text{CROP}_{\text{BR}, R + B(N + M), B(N + M)}  \mathcal{F}^{-1} \big\{ \mathcal{F} \big\{ \vb{E}_{\text{n2f}} \big\} \odot \mathcal{F} \big\{ \text{PAD}_{\text{TL},R, R + B(N + M)} (\vb{t}(f, \vb{p}) \odot \vb{E}_{\text{incident}}) \big\} \big\} \big|^2  \bigg)
    \label{differentialG3}
\end{aligned}
\end{equation}
where $\vb{E}_{\text{n2f} }$ is the Green's function of the near-to-far-field transformation, $\text{BIN}_B$ is an operator that ``bins" a 2D grid by integrating $B \times B$ adjacent values, and the absolute value is applied elementwise to the components of the vector. Equation~\eqref{differentialG3} depends on $\vb{p}$ only through the surrogate model $\vb{t}(f, \vb{p})$. The final step is to propagate the differential to the surrogate and express the right hand side as the dot product of a vector with $\partial \vb{p}$. One can then read off $ \partial (\vb{x}^{\mathsf{T}} \hat{G} (f, \vb{p})^{\mathsf{T}} \vb{y}) / \partial \vb{p}$ from the expression. Carrying out the algebra yields
\begin{equation}
\begin{aligned}
\partial \big( \vb{x}^{\mathsf{T}} \hat{G}(f, \vb{p})^{\mathsf{T}} \vb{y} \big) = 2 \Re & \bigg[ \frac{\partial \vb{t}(f, \vb{p})}{\partial \vb{p}} 
\text{diag} \{ \vb{E}_{\text{incident}}\} \text{CROP}_{\text{TL}, R + B(N + M), R} \\ 
&\mathcal{F}\big\{ \mathcal{F}^{-1}\{ \text{CROP}^{\mathsf{T}}_{\text{BR}, R + B(N + M), B(N + M)} \vb{z} \}  \odot \mathcal{F}\{ \vb{E}_{\text{n2f}}\} \big\} 
\bigg]^{\mathsf{T}} \cdot \partial \vb{p}
\end{aligned}
\end{equation}
where $\text{diag} \{ \vb{E}_{\text{incident}}\}$ is an $R \times R$ diagonal matrix with $\vb{E}_{\text{incident}}$ on the diagonals, and where we have defined an auxiliary variable $\vb{z}$ given by
\begin{equation}
\vb{z} = \big( \text{BIN}_{B}^{\mathsf{T}} \mathcal{F} \{ \text{PAD}_{\text{TL},N,N+M} \vb{x} \} \odot  \mathcal{F}^{-1} \{ \text{PAD}_{\text{BR},M,N+M}\vb{y} \} \big) \odot \big( \text{FF}(\vb{t}(f, \vb{p}) \odot \vb{E}_{\text{incident}}) \big),
\end{equation}
where $\text{BIN}_{B}^{\mathsf{T}}$ is an operator that ``un-bins" a 2D grid by repeating each of its elements $B \times B$ times.

% References
\medskip

% Use the following code if you wish to generate your bibliography with BibTeX;
% replace the string "MSP-template" below with the name(s) of
% the BibTeX data base(s) you want to use.
% The resulting bibliography-output (the content of the .bbl file)
% must be pasted back into this file before submission.
% Please also include your BibTeX data base file(s) in your submission
% so that we can re-run BibTeX if necessary.
%
\bibliographystyle{MSP}
\bibliography{bibliography}

% Figures/tables and captions
% Permission statements are required for all figures reproduced or adapted from previously published articles/sources. Please also ensure that all necessary permissions to reproduce images have been received
% Please remove these statements for original figures

\end{document}